\documentclass[
prl,preprint,
superscriptaddress,
showpacs,
amsmath,amssymb,
aps
]{revtex4-1}

\usepackage{graphicx}% Include figure files
\usepackage{amsmath}
\usepackage[version=3]{mhchem}
\usepackage[usenames]{color}
\usepackage{braket}
\usepackage{color}
\usepackage{braket}
\usepackage[colorlinks=true, linkcolor=blue, citecolor=blue, urlcolor=blue,pdftex]{hyperref}
\definecolor{newcolor}{rgb}{0.9,0,0.1}

\newcommand{\figref}[1]{Fig.~\ref{#1}}
\newcommand{\unit}[2]{$#1\,\text{#2}$}
\newcommand{\equnit}[3]{$#1=#2\,\text{#3}$}

\begin{document}

\title{Large spin-orbit splitting of deep in-gap defect states of engineered sulfur vacancies in monolayer \ce{WS2}}

\author{Bruno Schuler}
\email[]{bschuler@lbl.gov}
\affiliation{Molecular Foundry, Lawrence Berkeley National Laboratory, California 94720, USA}
\author{Diana Y. Qiu}
\affiliation{Department of Physics, University of California at Berkeley, Berkeley, California 94720, USA}
\affiliation{Materials Sciences Division, Lawrence Berkeley National Laboratory, Berkeley, California 94720, USA}

\author{Sivan Refaely-Abramson}
\affiliation{Molecular Foundry, Lawrence Berkeley National Laboratory, California 94720, USA}
\affiliation{Department of Physics, University of California at Berkeley, Berkeley, California 94720, USA}

\author{Christoph Kastl}
\author{Christopher T. Chen}
\affiliation{Molecular Foundry, Lawrence Berkeley National Laboratory, California 94720, USA}

\author{Sara Barja}
\affiliation{Molecular Foundry, Lawrence Berkeley National Laboratory, California 94720, USA}
\affiliation{Departamento de F\'isica de Materiales, Centro de F\'isica de Materiales, University of the Basque Country UPV/EHU-CSIC, Donostia-San Sebasti\'an 20018, Spain}
\affiliation{Ikerbasque, Basque Foundation for Science}
\affiliation{Donostia International Physics Center, Donostia-San Sebastián 20018, Spain}

\author{Roland J. Koch}
\affiliation{Advanced Light Source, Lawrence Berkeley National Laboratory, California 94720, USA}
\author{D. Frank Ogletree}
\author{Shaul Aloni}
\author{Adam M. Schwartzberg}
\affiliation{Molecular Foundry, Lawrence Berkeley National Laboratory, California 94720, USA}

\author{Jeffrey B. Neaton}
\email[]{jbneaton@lbl.gov }
\affiliation{Molecular Foundry, Lawrence Berkeley National Laboratory, California 94720, USA}
\affiliation{Department of Physics, University of California at Berkeley, Berkeley, California 94720, USA}
\affiliation{Kavli Energy Nanoscience Institute at Berkeley, Berkeley, California 94720, USA}

\author{Steven G. Louie}
\email[]{sglouie@berkeley.edu}
\affiliation{Department of Physics, University of California at Berkeley, Berkeley, California 94720, USA}
\affiliation{Materials Sciences Division, Lawrence Berkeley National Laboratory, Berkeley, California 94720, USA}

\author{Alexander Weber-Bargioni}
\email[]{afweber-bargioni@lbl.gov}
\affiliation{Molecular Foundry, Lawrence Berkeley National Laboratory, California 94720, USA}

\begin{abstract}
Structural defects in 2D materials offer an effective way to engineer new material functionalities beyond conventional doping. Here, we report the direct experimental correlation of the atomic and electronic structure of a sulfur vacancy in monolayer \ce{WS2} by a combination of CO-tip noncontact atomic force microscopy and scanning tunneling microscopy. 
Sulfur vacancies, which are absent in as-grown samples, were deliberately created by annealing in vacuum. Two energetically narrow unoccupied defect states followed by vibronic sidebands provide a unique fingerprint of this defect. Direct imaging of the defect orbitals reveals that the large splitting of $252\pm 4$\,meV between these defect states is induced by spin-orbit coupling.
\vspace{2cm}
\end{abstract}

%\date{\today}
\maketitle

Transition metal dichalcogenides (TMDs) and other layered materials have recently attracted considerable interest because of their unique properties arising from the combination of quantum confinement, reduced screening, strong spin-orbit coupling and lack of inversion symmetry in the monolayer limit~\cite{manzeli20172d}. 
The strong confinement, however, also causes TMD properties to be particularly sensitive to defects~\cite{lin2016defect}.
Structural defects in TMDs are thought to substantially modify optoelectronic properties and induce catalytic functionality to the otherwise inert surface. 
Particularly, chalcogen vacancies have been attributed to a variety of phenomena including single-photon emission~\cite{srivastava2015optically}, defect-bound excitons~\cite{tongay2013defects,chow2015defect,carozo2017optical,zhang2017defect}, catalytic activity~\cite{li2016activating} and hopping transport~\cite{qiu2013hopping}. In most of these studies, the chalcogen vacancy functionality was only indirectly inferred by the presence of this defect in transmission electron microscopy (TEM)~\cite{komsa2012two,qiu2013hopping,zhou2013intrinsic,komsa2015native,hong2015exploring,wang2018atomic}. Moreover, TMD monolayers are known to be electron beam sensitive and vacancy defects can be created in-situ by knock-on or radiolysis effects~\cite{komsa2012two,qiu2013hopping,zhou2013intrinsic,komsa2015native,hong2015exploring,wang2018atomic}, as reflected in the high reported defect densities on the order of $10^{13}$\,cm$^{-2}$, even in exfoliated samples,~\cite{hong2015exploring} with an estimated vacancy generation rate of about $5\times10^{10}$\,cm$^{-2}$s$^{-1}$~ \cite{komsa2012two,qiu2013hopping}. \\

A decisive factor for the functionality of defects is the creation of defect states in the band gap of the semiconductor.
While TEM can routinely resolve the atomic lattice, the electronic structure around the Fermi level is not easily accessible by TEM.
Conversely, scanning tunneling microscopy (STM) can probe the electronic structure of single defects~\cite{lu2014bandgap,hildebrand2014doping,peng2015molecular,liu2016point,vancso2016intrinsic,zhang2017defect,barja2017correlating,lin2018realizing}. Yet, the defect assignment is not straightforward because their STM contrast is dominated by their electronic states, and tip-dependent contrast inversion makes it difficult to assign lattice sites. Both of these complications have lead to recent contradictory defect identification in TMDs by STM~\cite{liu2016point,zhang2017defect,lin2018realizing}.\\

In this study, we use a combination of low-temperature scanning tunneling microscopy/ spectroscopy (STM/STS), CO-tip noncontact atomic force microscopy (nc-AFM) and \textit{ab initio} GW calculations to unambiguously identify and characterize the chalcogen vacancy in \ce{WS2}.
We find that chalcogen vacancies are largely absent in as-grown TMD samples under ambient conditions. Chalcogen vacancies were, however, deliberately created by \textit{in vacuo} annealing at elevated temperatures. In STS, the sulfur vacancy in WS$_2$ exhibits a characteristic fingerprint with two narrow unoccupied defect states accompanied by vibronic satellite peaks. The observed splitting between the two defect peaks is caused by extraordinarily strong spin-orbit coupling. This effect has not been observed experimentally before. In light of this new evidence, perceptions of the abundance and functionality of ``the most discussed defect in TMDs'' need to be revisited. It also opens up new avenues for defect engineering in the context of valleytronics, solitary dopant optoelectronics and catalysis.\\

The \ce{WS2} samples are grown using chemical vapor deposition (CVD) on graphitized SiC substrates (see Supplemental Material~\cite{SI} for further details)~\cite{Kastl2017cvd}. As argued in a recent paper~\cite{barja2017correlating}, chalcogen site defects are abundant but they can be identified as oxygen substituents rather than chalcogen vacancies with a radically different electronic structure~\cite{lu2015atomic,barja2017correlating}. Undecorated sulfur vacancies can, however, be generated by annealing or ion bombardment in vacuum as reported previously~\cite{komsa2012two,tongay2013defects,klein2017robust,liu2017temperature}. Calculations also showed that in vacuum, the chalcogen vacancy has the lowest formation energy of any intrinsic defect in several TMD materials~\cite{zhou2013intrinsic,hong2015exploring}. Note that this is not necessarily the case when other molecules are present~\cite{barja2017correlating}. 
\\

\begin{figure}[h]
\includegraphics[width=0.5\textwidth]{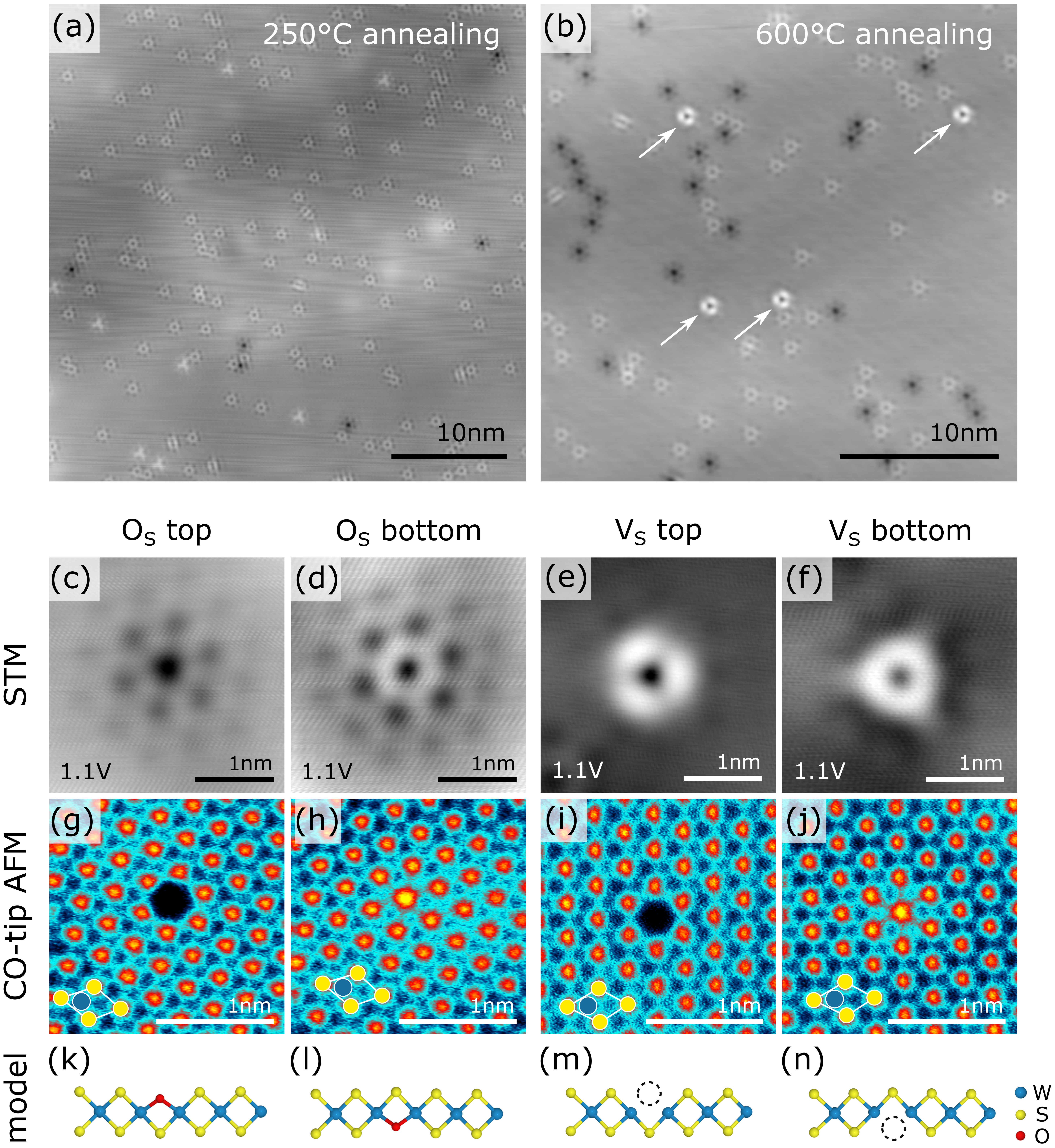}
\caption{\label{fig:AFM}
\textbf{Sulfur vacancy and O substituent.} 
 (a,b) STM topography (\equnit{I}{20}{pA},\equnit{V}{1.1}{V}) of CVD-grown monolayer \ce{WS2}. (a) After low temperature sample annealing ($\approx 250\,^\circ$C) no chalcogen vacancy defects were observed. Oxygen substituents at a sulfur sites are most abundant. (b) By in-vacuum annealing at about 600\,$^\circ$C, sulfur vacancies can be generated (white arrows).
(c-f) STM topography (\equnit{I}{20}{pA},\equnit{V}{1.1}{V}) of a oxygen substituting sulfur (O$_\text{S}$)  in the top (c) and bottom (d) sulfur plane as well as a sulfur vacancy in the top (e) and bottom (f) sulfur plane. (g-j) Corresponding CO-tip nc-AFM images of the same defects as in (c-f). The unit cell has been indicated as a guide to the eye. Yellow: S atom, blue: W atom. (k-n) DFT calculated defect geometry.
}
\end{figure}

In \figref{fig:AFM}(a,b) large-scale STM images of monolayer \ce{WS2} after annealing at 250$\,^\circ$C (a) and 600$\,^\circ$C (b) in ultrahigh vacuum are shown. After low temperature annealing, no sulfur vacancy defects are observed. Instead, we mainly find oxygen substituting sulfur in the top and bottom sulfur sublattice (O$_\text{S}$ top and O$_\text{S}$ bottom)~\cite{barja2017correlating}, and tungsten substitutions. We collectively refer to these point defects as 'as-grown' defects. After the high temperature anneal [\figref{fig:AFM}(b)], sulfur vacancies in the top and bottom sulfur sublattice (V$_\text{S}$ top and V$_\text{S}$ bottom) are observed along with all as-gown \ce{WS2} point defects. \figref{fig:AFM}(c-j) shows STM and nc-AFM maps of a substituted oxygen defect and a sulfur vacancy in both the top and bottom sulfur layer (facing the tip and the underlying graphene, respectively). Substitutional oxygen and sulfur vacancies can be clearly distinguished from each other in STM [c.f. \figref{fig:AFM}(c,d) and \figref{fig:AFM}(e,f)]. Note that a similar STM contrast as depicted in \figref{fig:AFM}(c,d), which we assign to O$_\text{S}$, has been reported for \ce{MoS2} and \ce{WSe2}, and was ascribed either to a sulfur vacancy~\cite{liu2016point} or a tungsten vacancy~\cite{zhang2017defect,lin2018realizing}. Such inconsistencies already reveal the difficulty of defect structure identification based on STM alone.
\\

Our defect structure assignment is founded on the CO-tip nc-AFM images, that are in excellent agreement with simulations based on the probe particle model~\cite{hapala2014mechanism} (see Fig.~S2~\cite{SI}), as well as the distinct defect electronic structure as discussed in detail below. 
While the CO-tip in nc-AFM is exceptionally sensitive to the outermost surface layer, it is difficult to distinguish between vacancies and oxygen substituents, which are located slightly below the surface sulfur plane. For both the oxygen substituent and the sulfur vacancy the neighboring surface sulfur atoms relax similarly. When the defect is located on the top sulfur layer, both defects appear as a missing sulfur atom [c.f. \figref{fig:AFM}(g,i)], because the oxygen atom of O$_\text{S}$ binds closer to the tungsten plane. In turn if the defect is in the bottom sulfur layer both V$_\text{S}$ and O$_\text{S}$ appear as a S atom that is protruding from the surface [c.f. \figref{fig:AFM}(h,j)]. In direct comparison (i.e. when measured in the same image), O$_\text{S}$ appears slightly more attractive than V$_\text{S}$ (see Fig.~S1~\cite{SI}). We would like to point out that it would be difficult to discriminate an oxygen substituent from a chalcogen vacancy by TEM because the sulfur atom on the opposite side of the layer masks the presence of the low atomic number O atom. 
Electronically, however, the sulfur vacancy and the oxygen substituent are fundamentally different. While O$_\text{S}$ does not feature defect states in the \ce{WS2} band gap~\cite{barja2017correlating}, V$_\text{S}$ does have pronounced deep in-gap defect states, which will be discussed next.
\\

\begin{figure*}[h]
\includegraphics[width=0.8\textwidth]{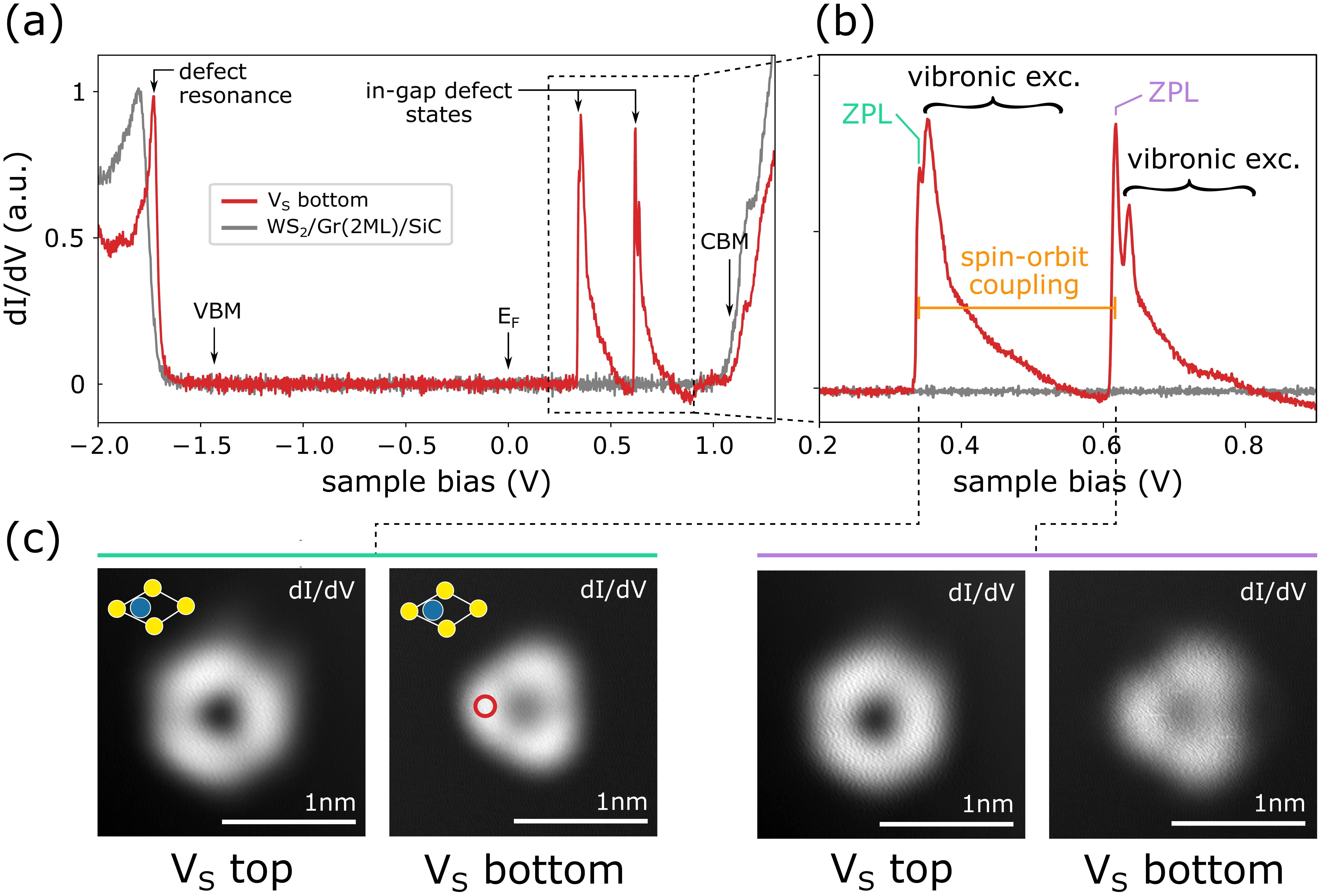}
\caption{\label{fig:orbitals}
\textbf{Sulfur vacancy defect states.} (a) STS spectra recorded on a sulfur bottom vacancy and the pristine \ce{WS2} monolayer on bilayer graphene. The valence band maximum (VBM), conduction band minimum (CBM), the Fermi level (E$_\text{F}$), the filled-states defect resonance and the unoccupied in-gap defect states are indicated. The VBM at K is resolved at a closer tunneling distance~\cite{zhang2015probing}. The spectrum position is marked by the red circle in (c). (b) STS spectra of the deep unoccupied V$_\text{S}$ defect states. The two zero-phonon lines (ZPL) and the subsequent vibronic satellite peaks are labelled. The splitting between the ZPL peaks is due to spin-orbit coupling. (c) Constant-height dI/dV maps of the two V$_\text{S}$ defect states corresponding to the ZPL resonances of both the top and bottom V$_\text{S}$. The \ce{WS2} unit cell has been indicated (to scale). Yellow: S, blue: W.}
\end{figure*}

In \figref{fig:orbitals}(a,b), STS spectra of a single V$_\text{S}$ in the bottom sulfur layer are shown.
We find that both the top and bottom V$_\text{S}$ are electronically equivalent, indicating the negligible influence of the graphene substrate (see Fig.~S3~\cite{SI}).
The defect introduces a series of sharp resonances at positive sample bias (above the Fermi level) and a single resonance at negative bias (below the Fermi level). The occupied defect state resonance is located about \unit{300\pm 10}{meV} below the valence band maximum (VBM) overlapping with delocalized bulk states, similar to the defect resonance observed for O$_\text{S}$. Strikingly, we find two unoccupied defect states at \unit{774\pm5}{meV} and \unit{522\pm5}{meV} below the conduction band minimum (CBM), deep in the band gap. Each of these defect state resonances is accompanied by satellite peaks, which we assign to vibronic excitations due to inelastic tunneling electrons. Similar spectral features are commonly observed in tunneling spectroscopy of molecules~\cite{qiu2004vibronic}. We note that the vibronic features appear in dI/dV as opposed to d$^2$I/dV$^2$, because of the double-barrier tunneling junction geometry~\cite{repp2010coherent}.\\

We expect that several phonon modes are excited by the transient electron attachment during tunneling. This phonon emission results from a change in equilibrium geometry of different defect oxidation states~\cite{Repp2005a}. The relaxation is particularly large for polar solids and for defects with strongly localized wavefunctions, both of which apply to the S vacancy in \ce{WS2}. We estimated an effective electron-phonon coupling strength by employing the Franck-Condon model. We fitted the elastic and first inelastic peak of the dI/dV spectrum to this single-mode model. From the relative peak intensities and the vibronic peak separation we estimate a Huang-Rhys factor of $S = 1.2 \pm 0.1$ with a phonon mode of $\hbar\omega_0 = 12\pm 2\,$meV and broadening of $\Gamma = 10 \pm 2\,$meV (full width at half maximum) for the lower energy defect state and $S = 0.7\pm 0.1$ with $\hbar\omega_0 = 18\pm 2\,$meV and $\Gamma = 10\pm 2\,$meV for the higher energy defect state. These vibrational frequencies of the lowest excited phonon are in the range of the TA(M) (\unit{120}{cm$^{-1}$})~\cite{molina2011phonons} and ZA(M) (\unit{146.5}{cm$^{-1}$})~\cite{molas2017raman} acoustic phonons of monolayer \ce{WS2} as predicted by \textit{ab initio} theory and measured by low-temperature Raman spectroscopy, respectively. The vacancy is also expected to introduce additional quasi-local modes in the acoustic phonon spectrum~\cite{peng2016beyond}.
However, the simple Franck-Condon approximation underestimates the dI/dV signal strength and broadening observed at higher energies (see Fig.~S4). This is consistent with the coexcitation of higher energy phonon modes (and multiple quanta thereof).
\\

The V$_\text{S}$ defect's purely electronic states at the lowest excitation energy are denoted zero-phonon line (ZPL) in \figref{fig:orbitals}(b), following the conventions used in absorption/emission spectroscopy. Most notably, both defect resonances exhibit the same spatial electron distribution as evident from the dI/dV maps shown in \figref{fig:orbitals}(c). We attribute this observation to a lifted degeneracy induced by spin-orbit coupling (SOC). Hence, the energy separation between the two elastic excitations quantifies the spin-orbit interaction as \unit{252}{meV}, which is exceptionally large. As we will see below, each of the two peaks is composed of two degenerate defect states. In dI/dV maps, these states are imaged as a superposition of the charge densities of the degenerate orbitals.
The defect states appear different for the top and bottom V$_\text{S}$ since the defect and its orbital are not mirror-symmetric with respect to the tungsten plane.\\

\begin{figure*}[h]
\includegraphics[width=0.9\textwidth]{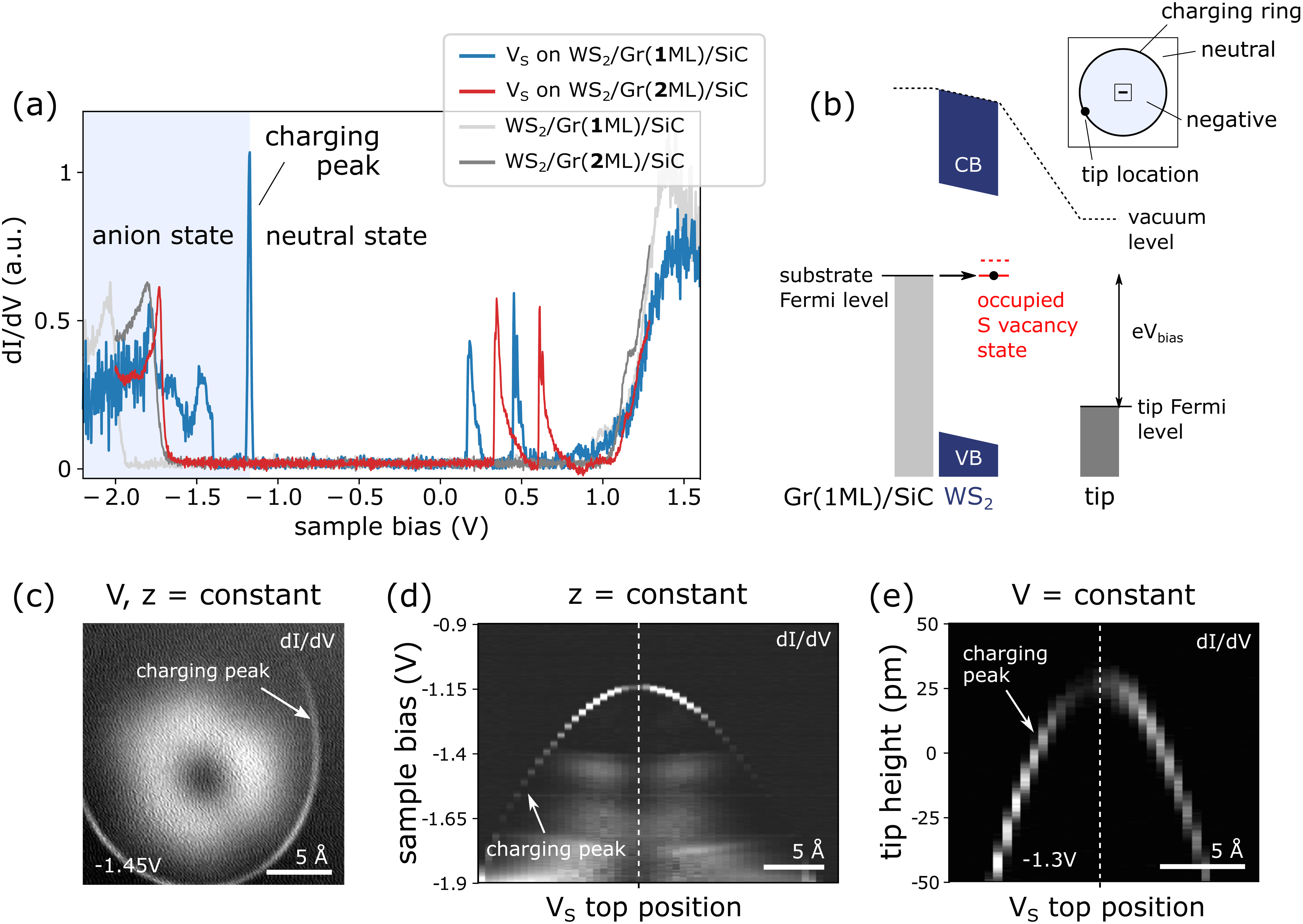}
\caption{\label{fig:charging}
\textbf{Tip-induced sulfur vacancy charging on Gr(1ML)/SiC substrates.} (a) STS spectra recorded on a V$_\text{S}$ on \ce{WS2}(1ML) on Gr(1ML)/SiC (blue) and on Gr(2ML)/SiC (red). The V$_\text{S}$ defect states on Gr(1ML)/SiC are at lower energies than on Gr(2ML)/SiC and a pronounced charging peak is observed at about \unit{-1.2}{V} along with an upwards shift of the valence band. \ce{WS2} on Gr(1ML)/SiC (light gray) has as a lower VBM than \ce{WS2} on Gr(2ML)/SiC (dark gray). (b) Schematic of the tip-induced band bending at a bias and tip position where the V$_\text{S}$ (red line)  becomes resonant with the Fermi level of the substrate (at the 'charging ring'). The V$_\text{S}$ state at zero field energy is indicated by the red dashed line. (c) dI/dV map of V$_\text{S}$ on Gr(1ML)/SiC at \unit{-1.45}{V}. The charging peak is visible as a circular feature around the defect. Inside the ring the defect is charged, outside it is neutral. (d) STS spectra at negative sample bias recorded across the V$_\text{S}$. (e) dI/dV intensity at \unit{-1.3}{V} recorded as a function of tip height across the defect.
}
\end{figure*}

The spectra in \figref{fig:orbitals}(a,b) are measured on monolayer \ce{WS2} on bilayer graphene on silicon carbide [\ce{WS2}(1ML)/Gr(2ML)/SiC]. On a monolayer graphene substrate, the filled-state spectrum is qualitatively different. We find an additional major resonance around \unit{-1.2}{V} and the valence band is pushed upwards [see \figref{fig:charging}(a)]. 
We attribute this additional feature to the stationary tip-induced charging of the defect (commonly referred to as charging peak~\cite{teichmann2008controlled}). The charging peak is identified by its energetic shift towards more negative biases for farther tip-defect distances (both laterally and vertically) as shown in \figref{fig:charging}(c-e).
At sufficiently high negative bias, tip-induced band bending pulls the lowest unoccupied defect state below the substrate Fermi level. Therefore the defect becomes negatively charged in the vicinity of the tip as illustrated in \figref{fig:charging}(b). 
If the tip is within tunneling distance of the defect state, the additional electron in the defect can drain to the tip. Therefore, the contrast of the dI/dV map \figref{fig:charging}(c) within the charging ring corresponds the defect state that is occupied by the field-induced charging, resembling the lowest formerly unoccupied V\textsubscript{S} orbital. On average the defect stays negatively charged because it is immediately refilled by electrons from the substrate.\\

The different behavior of mono- vs bilayer graphene substrates can be explained by the energetic shift of the V\textsubscript{S} defect states by \unit{160}{meV} towards lower energies. A similar shift is observed for the \ce{WS2} VBM, which can be attributed to the smaller screening of monolayer graphene that increases the band gap~\cite{ugeda2014giant} and a change in work function~\cite{mammadov2017work}. 
Using the charging peak, a 11\% voltage drop across the \ce{WS2} layer (at the chosen tunneling condition) was estimated.  All defect state energies stated previously were corrected for this effect.\\

\begin{figure*}[h]
\includegraphics[width=0.9\textwidth]{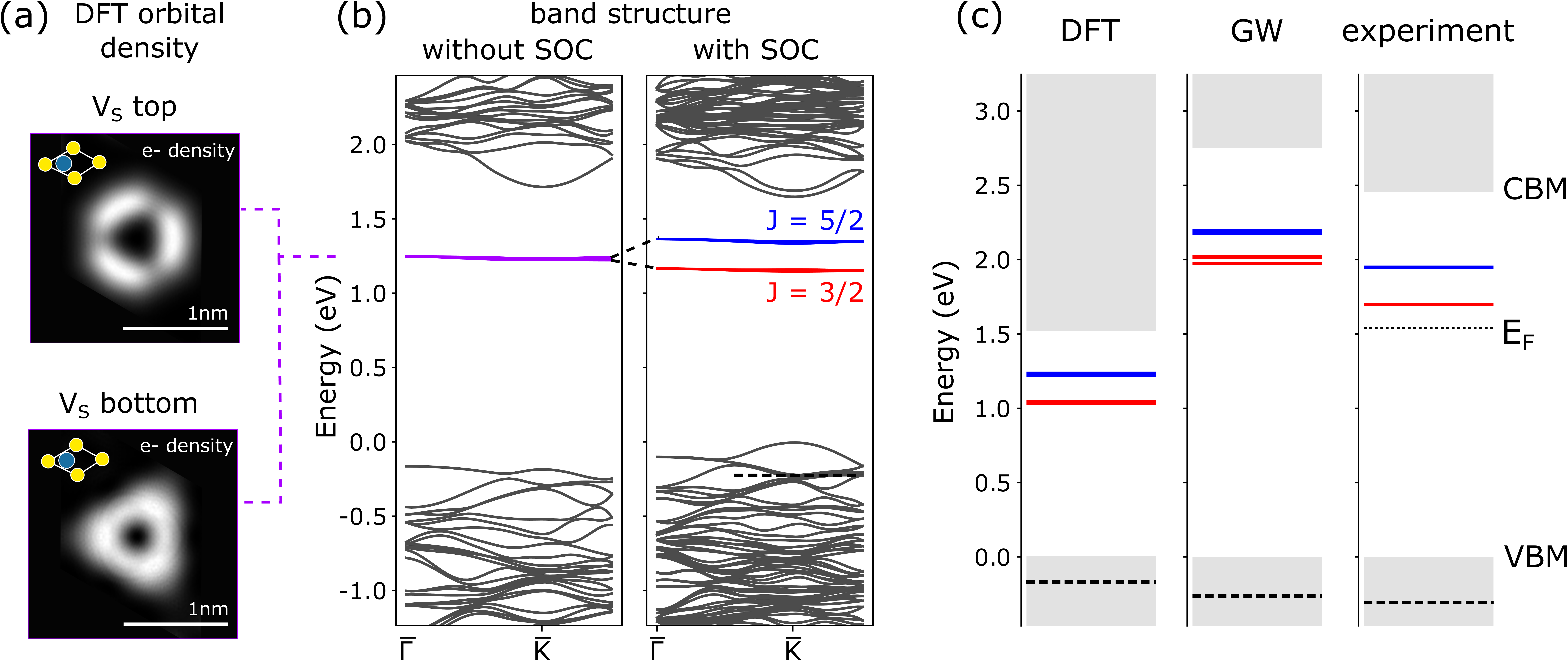}
\caption{\label{fig:SOC}
\textbf{Calculated DFT and GW defect levels including SOC.} All calculations were done in a 5$\times$5 supercell containing a single sulfur vacancy. (a) DFT-LDA level constant height slice of the orbital density of the in-gap defect state \unit{8}{\AA} above (corresponding to V$_\text{S}$ top) and \unit{8}{\AA} below (corresponding to V$_\text{S}$ bottom) the \ce{WS2} monolayer. The \ce{WS2} unit cell has been indicated (to scale). Yellow: S, blue: W. (b) Comparison of the band structure excluding and including SOC. The formerly degenerate defect state (purple) splits into two states (red and blue) with dominant contributions of $J=3/2$ and $J=5/2$ of the total angular momentum. The black dashed line indicates the occupied defect resonance overlapping the valence band. (c) Comparison of the defect state energies calculated on the DFT and GW level and the corresponding experimental values (on a monolayer graphene substrate). Screening effects by the substrate have not been considered in the calculations. The gray boxes represent the delocalized states of the \ce{WS2} layer with the conduction band minimum (CBM) and valence band maximum (VBM). The VBM of the DFT, GW calculations and the experiment has been aligned for comparability. $E_\text{F}$ denotes the experimental Fermi level.}
\end{figure*}

To verify our interpretation of the spin-orbit split defect states of the sulfur vacancy, we calculated the electronic structure of a WS$_2$ monolayer with chalcogen vacancy point defects using the \textit{ab initio} GW approach. We constructed a supercell consisting of 5 unit cells along each crystalline-axis direction of the monolayer plane (namely 50 S atoms and 25 W atoms) and then removed a single chalcogen atom~\cite{refaely2018defect}. We accounted for spin-orbit coupling via a fully-relativistic, noncollinear density functional theory (DFT) starting point as implemented in Quantum Espresso and built one-shot GW energy corrections on top of it, as implemented in the BerkeleyGW package (see Supplemental Material~\cite{SI} for full computational details). \figref{fig:SOC} shows the resulting DFT band structures computed within the local density approximation (LDA) and GW energy levels. At both the DFT and GW level, the sulfur vacancy introduces four unoccupied in-gap states, which form two pairs of nearly-degenerate flat bands in the gap, corresponding to the two deep in-gap resonances in the dI/dV. The nearly-degenerate states are time-reversal pairs whose degeneracy has been lifted slightly due to interaction between periodic images of the 5$\times$5 supercell. The charge distributions of the two in-gap states are highly localized around the S vacancy, as depicted in \figref{fig:SOC}(a). The calculated defect orbitals for the top and bottom V$_\text{S}$ are in excellent agreement with the corresponding dI/dV maps in \figref{fig:orbitals}(c). In addition, an occupied doubly-degenerate defect-localized resonant state appears in the valence band region, in agreement with experimental observation [black dashed line in \figref{fig:SOC}(b,c)]. \figref{fig:SOC}(c) shows the resulting energy gaps. The one-shot GW correction opens the VBM-CBM gap to 2.8\,eV (compared to the calculated DFT gap of 1.6\,eV), comparing well with the experimental value of 2.5\,eV.  The error bar of the bandgap within our calculation is estimated to be 150\,meV - as a result of the sensitivity of the GW approach to the DFT starting point and to structural effects. Note that screening effects from the graphene substrate are not included in the calculations but can be expected to reduce the gap by a few hundred meV~\cite{ugeda2014giant,Bradley2015,naik2018substrate}. Importantly, the GW gap between the highest in-gap defect state and the conduction band is 0.6\,eV, in reasonable agreement with the measured value of 0.52\,eV in experiment. The spin-orbit energy splitting between the two doubly-degenerate in-gap states is 180\,meV from our calculations.\\

Our theoretical calculations also shed light on the spin-orbit splitting and character of the in-gap states. Although many theoretical studies predicted the chalcogen vacancy to form in-gap states~\cite{tongay2013defects,zhou2013intrinsic,komsa2015native,hong2015exploring,vancso2016intrinsic,naik2018substrate}, only a few explicitly consider the effect of SOC~\cite{yuan2014effect,li2016strong,khan2017electronic}.
We find that each in-gap peak in the dI/dV spectrum corresponds to two degenerate states belonging to a time-reversal pair. The character of the in-gap states consists primarily of W $d$-states, which are responsible for the large magnitude of the spin-orbit splitting, with some smaller contributions from the S 3$p$ and W 5$p$ states (see Supplemental Material~\cite{SI}). The lower-energy in-gap states have a larger contribution from J = 3/2 states, and the higher energy in-gap states have a larger contribution from J = 5/2 states. The close correspondence between the in-gap states in theory and experiment are a clear indication of the presence of sulfur vacancies. Importantly, the hybridization between defect-localized in-gap states and delocalized, pristine-like states can lead to significant valley depolarization~\cite{refaely2018defect} and suggests a path to control spin-valley selectivity through defect engineering.\\

In summary, we created and identified individual sulfur vacancies in monolayer \ce{WS2} by a combination of atomic-resolution nc-AFM, STS and \textit{ab initio} GW calculations. We show that a sulfur vacancy gives rise to two unoccupied in-gap defect states that appear as sharp resonances followed by vibronic satellite peaks in STS. The deep in-gap states act as a strong atom trap, which explains why undecorated chalcogen vacancies are largely absent in as-grown TMD samples under ambient conditions.
Remarkably, the degeneracy between the four V$_\text{S}$ defect orbitals is lifted by spin-orbit interaction into two pairs of degenerate orbitals as revealed by direct STM orbital imaging and state-of-the-art DFT and GW calculations.
The exceptionally large spin-orbit splitting between the sulfur vacancy states was measured to be \unit{252}{meV}, consistent with our theoretical predictions. 
These results suggest that the controllable introduction of chalcogen vacancies in vacuum could be used to tune the spin-valley polarization in TMDs and potentially to induce single-photon emission. Moreover, we speculate that the reactive vacancy sites could be used to trap diffusing adatoms, offering the possibility to embed arbitrary dopants into the 2D TMD matrix. This concept could be particularly interesting to study the interaction of magnetic impurities in a highly-correlated material or to catalytically activate the inert basal plane of TMDs.

\bibliography{Svac}

\section*{Acknowledgments}
We thank Andreas Schmid, Katherine Cochrane and Nicholas Borys for discussions. B.S. appreciates support from the Swiss National Science Foundation under project number P2SKP2\_171770.
Theoretical work was supported by the Center for Computational Study of Excited State Phenomena in Energy Materials, which is funded by the U.S. Department of Energy, Office of Science, Basic Energy Sciences, Materials Sciences and Engineering Division under Contract No. DE-AC02-05CH11231, as part of the Computational Materials Sciences Program. Work performed at the Molecular Foundry was also supported by the Office of Science, Office of Basic Energy Sciences, of the U.S. Department of Energy under the same contract number. S.R.A acknowledges Rothschild and Fulbright fellowships. S.B. acknowledges support by the European Union under FP7-PEOPLE-2012-IOF-327581 and Spanish MINECO (MAT2017-88377-C2-1-R). This research used resources of the National Energy Research Scientific Computing Center (NERSC), a DOE Office of Science User Facility supported by the Office of Science of the U.S. Department of Energy under Contract No. DE-AC02-05CH11231.\\

\clearpage

\section*{Figures}

\end{document}